\documentclass[aps,prl,reprint,groupedaddress,showpacs,amsfonts,amsmath,amssymb,superscriptaddress]{revtex4-1}
\usepackage{graphicx}
\usepackage{bm}
\usepackage{color}
\usepackage[colorlinks,urlcolor=blue,linkcolor=blue,anchorcolor=blue,citecolor=blue, bookmarks]{hyperref}

\begin{document}

\title{Realization of metallic state in 1\emph{T}-TaS$_{2}$ with persisting long-range order of charge density wave}

\author{Xin-Yang Zhu}
\affiliation{National Laboratory of Solid State Microstructures and School of Physics, Nanjing University, Nanjing 210093, China}
\author{Shi Wang}
\affiliation{National Laboratory of Solid State Microstructures and School of Physics, Nanjing University, Nanjing 210093, China}
\author{Zhen-Yu Jia}
\affiliation{National Laboratory of Solid State Microstructures and School of Physics, Nanjing University, Nanjing 210093, China}
\author{Li Zhu}
\affiliation{National Laboratory of Solid State Microstructures and School of Physics, Nanjing University, Nanjing 210093, China}
\author{Qi-Yuan Li}
\affiliation{National Laboratory of Solid State Microstructures and School of Physics, Nanjing University, Nanjing 210093, China}
\author{Wei-Min Zhao}
\affiliation{National Laboratory of Solid State Microstructures and School of Physics, Nanjing University, Nanjing 210093, China}
\author{Cheng-Long Xue}
\affiliation{National Laboratory of Solid State Microstructures and School of Physics, Nanjing University, Nanjing 210093, China}
\author{Yong-Jie Xu}
\affiliation{National Laboratory of Solid State Microstructures and School of Physics, Nanjing University, Nanjing 210093, China}
\author{Zhen Ma}
\affiliation{National Laboratory of Solid State Microstructures and School of Physics, Nanjing University, Nanjing 210093, China}
\author{Jinsheng Wen}
\affiliation{National Laboratory of Solid State Microstructures and School of Physics, Nanjing University, Nanjing 210093, China}
\affiliation{Collaborative Innovation Center of Advanced Microstructures, Nanjing University, Nanjing 210093, China}
\author{Shun-Li Yu}
\email{slyu@nju.edu.cn}
\affiliation{National Laboratory of Solid State Microstructures and School of Physics, Nanjing University, Nanjing 210093, China}
\affiliation{Collaborative Innovation Center of Advanced Microstructures, Nanjing University, Nanjing 210093, China}
\author{Jian-Xin Li}
\email{jxli@nju.edu.cn}
\affiliation{National Laboratory of Solid State Microstructures and School of Physics, Nanjing University, Nanjing 210093, China}
\affiliation{Collaborative Innovation Center of Advanced Microstructures, Nanjing University, Nanjing 210093, China}
\author{Shao-Chun Li}
\email{scli@nju.edu.cn}
\affiliation{National Laboratory of Solid State Microstructures and School of Physics, Nanjing University, Nanjing 210093, China}
\affiliation{Collaborative Innovation Center of Advanced Microstructures, Nanjing University, Nanjing 210093, China}

\date{\today}

\begin{abstract}
Metallization of $1$\emph{T}-TaS$_{2}$ is generally initiated at the domain boundary of charge density wave (CDW), at the expense of its long-range order. However, we demonstrate in this study that the metallization of $1$\emph{T}-TaS$_{2}$ can be also realized without breaking the long-range CDW order upon surface alkali doping. By using scanning tunneling microscopy, we find the long-range CDW order is always persisting, and the metallization is instead associated with additional in-gap excitations. Interestingly, the in-gap excitation is near the top of the lower Hubbard band, in contrast to a conventional electron-doped Mott insulator where it is beneath the upper Hubbard band. In combination with the numerical calculations, we suggest that the appearance of the in-gap excitations near the lower Hubbard band is mainly due to the effectively reduced on-site Coulomb energy by the adsorbed alkali ions.
\end{abstract}

\maketitle

$1$\emph{T}-TaS$_{2}$ has been extensively studied for decades, as a unique transition metal dichalcogenide compound hosting the ground state of a Mott insulator \cite{J.Phys.Colloq.37.C4-165,Philos.Mag.B.39.229,Physica.B+C.99.183} intertwined with a commensurate charge density wave (CDW) \cite{PhysRevB.38.10734}. A series of temperature dependent CDW orders have been discovered \cite{Philos.Mag.B.39.229,PhysRevB.12.2220,Philos.Mag.31.255}, and a large electron-phonon coupling has been demonstrated in $1$\emph{T}-TaS$_{2}$ which may be associated with the formation of the CDW state \cite{BENEDEK19941,BENEDEK1994185}.
Realization of superconductivity in $1$\emph{T}-TaS$_{2}$, particularly by doping \cite{APL.102.192602,EPL.97.67005} or applying pressure \cite{nmat2318}, further makes it an alternate model system to study the interplay between superconductivity and strong correlation physics \cite{PhysRevB.94.125126}. Recently, $1$\emph{T}-TaS$_{2}$ attracts more attention because of the theoretical prediction and experimental exploration of the possible quantum spin liquid state \cite{Anderson1973.153,nphys4212,Law6996,npj.2.42,PhysRevB.96.195131,PhysRevB.96.081111}.

In general, the band filling $n$, the on-site Coulomb energy $U$ and the one-electron bandwidth $W$ together drive the Mott insulator transition in a coordinated manner \cite{RevModPhys.70.1039}. Unlike other conventional Mott insulators, the on-site Coulomb energy $U$ in $1$\emph{T}-TaS$_{2}$ is not large enough to directly open a Mott gap.
The essential ingredient here is that the presence of the commensurate CDW with a David star structure isolates a narrow energy band near the Fermi energy \cite{Physica.B+C.99.51}. This band mainly consists of the $5d$ orbitals at the center of the David star \cite{PhysRevB.73.073106}. Therefore, a moderate $U$ can open a Mott gap in this narrow band
which is half-filled.

The metallization of $1$\emph{T}-TaS$_{2}$ is believed to be accompanied with the breaking of the long-range CDW order, i.e., the metallic phase starts to initiate at the domain boundaries of the CDW state \cite{PhysRevLett.122.036802}, where the long-range order is simultaneously destructed. The metallization has be realized experimentally via applying pressure \cite{nmat2318},
chemical substitution \cite{PhysRevX.7.041054}, ultra-fast laser pulses \cite{Stojchevska177}, current excitation \cite{ncomms11442,Yoshidae1500606} or applying a pulse voltage \cite{ncomms10956}, which induce domain boundaries and suppress the long-range CDW order \cite{nmat2318,PhysRevX.7.041054,Stojchevska177,ncomms11442,Yoshidae1500606,ncomms10956}.  However, the nature of Mott insulator-metal transition in $1$\emph{T}-TaS$_{2}$ still remains elusive, partially owing to the complex correlation between the Mott insulator and CDW state. It is not even clear whether the collapse of the Mott gap is necessarily connected with the destruction of the long-range CDW order.

In this work, we explore the Mott insulator-metal transition of $1$\emph{T}-TaS$_{2}$ upon surface doping by alkali K atoms. The K atoms deposited onto the surface will transfer electrons to the host lattice to increase the band filling \cite{RAMIREZ2001424}, and the left K$^{+}$ ions effectively modify the crucial parameters $U/W$ of the Mott insulator-metal transition. At the meantime, such a method induces negligible orbital effect which may complicate the related physics \cite{PhysRevB.50.8816}.
By using scanning tunneling microscopy/spectroscopy (STM/STS), we directly characterize the electronic state evolution from Mott insulator to metal. Surprisingly, we find the long range CDW order is always persisting during the metallization, indicating that such Mott insulator-metal transition is fundamentally different from that triggered by the domain boundaries of the CDW \cite{PhysRevX.7.041054,PhysRevLett.95.126403,PhysRevLett.109.176403,PhysRevLett.112.206402}. The metallization is found to arise from the filling up of the Mott gap by additional in-gap excitations induced by the surface alkali doping. Though this phenomenon is common for conventional Mott insulators, such as the doped high-$T_c$ cuprates \cite{Nat.hys.12.1047}, we observe that the in-gap excitation is unexpectedly located above the lower Hubbard band (LHB), which is underneath the upper Hubbard band (UHB) in the case of electron doped conventional Mott insulators. Considering that the Mott insulating state in $1$\emph{T}-TaS$_{2}$ is derived from the isolated narrow band due to the CDW, we construct a single-band Hubbard model with a site-dependent $U$ to account for the possible effect of the surface K$^{+}$ ions. Using the cluster perturbation theory (CPT), we calculate the spectral function of electrons and find the qualitative consistence with experiments.
We therefore suggest that the K$^{+}$ ions adsorbed at the center of David star would effectively reduce the Coulomb energy $U$ and results in the appearance of the in-gap excitations near the LHB.

The $1$\emph{T}-TaS$_{2}$ single crystal was grown by chemical vapor transport method. STM and STS measurements were carried out in a low-temperature scanning tunneling microscope (Unisoku Co.) at $\sim$4.2 K in ultrahigh vacuum (UHV), with the base pressure of $1\times10^{-10}$ mbar. The clean surface of $1$\emph{T}-TaS$_{2}$ was achieved by cleaving the sample \emph{in-situ} in UHV. The sample was kept at $\sim$4.2 K and then quickly transferred to the evaporation chamber prior to K deposition. The K source (SAES, Alkali metal dispenser) was heated to $\sim$650 $^{\circ}$C, and the atomic flux of K is estimated to be $\sim$0.005 monolayer (ML)/s. After the deposition, the sample was transferred to the STM stage for scan without annealing. Due to the strong enough bonding to the $1$\emph{T}-TaS$_{2}$ surface, the K atoms are static during normal scan. STM topographic images were collected under the constant current mode. STS dI/dV spectra were taken with a lock-in amplifier and a typical ac modulation of $10$ mV and $879$ Hz.

\begin{figure}
  \centering
  \includegraphics[scale=1.0]{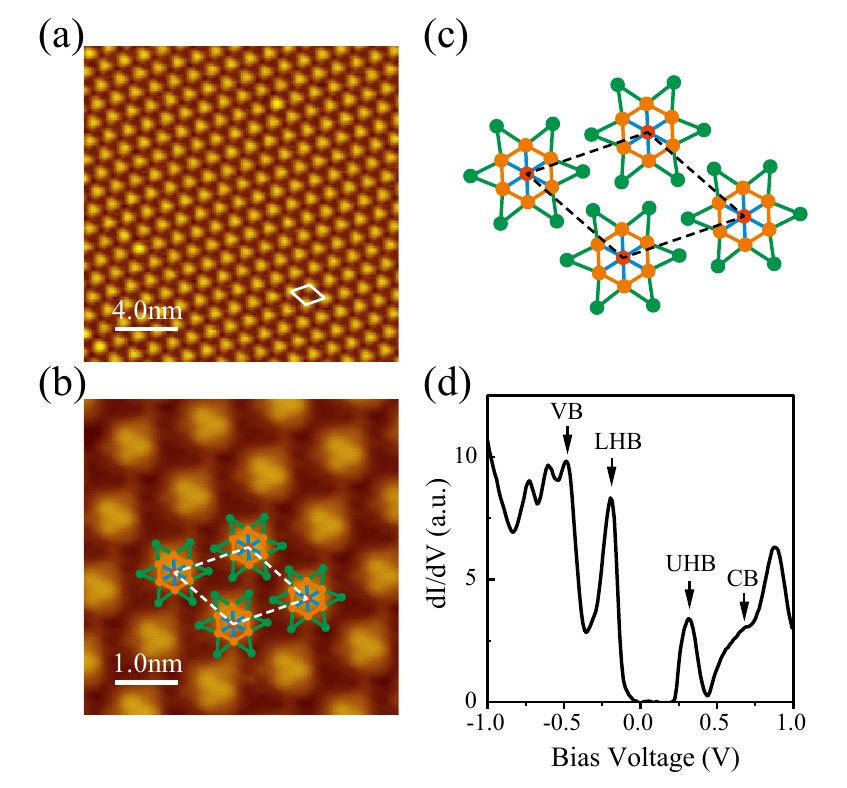}
  \caption{\label{david-star}(a) STM topographic image ($40 \times40$ $\mathrm{nm}^{2}$) taken at $\sim$4.2 K on the pristine $1$\emph{T}-TaS$_{2}$ surface, showing the $\sqrt{13}\times\sqrt{13}$ periodicity. $U=+1.0~\mathrm{V}$ and $I_{t}=100~\mathrm{pA}$. (b) High-resolution STM image ($5 \times5$ $\mathrm{nm}^{2}$) showing the atomic structure of the David star. The David star model is overlaid on the STM image. $U=+100~\mathrm{mV}$ and $I_{t}=200~\mathrm{pA}$. Individual S atoms are resolved as small blobs in each David star cluster, which enables us to identify the position of Ta atoms underneath. (c) Schematic of the top view David star structure. Different colors represent the inequivalent Ta atoms derived from different positions in the David star cluster. (d) Spatially-averaged dI/dV spectrum acquired with $U=+500~\mathrm{mV}$ and $I_{t}=100~\mathrm{pA}$.}
\end{figure}
The bulk $1$\emph{T}-TaS$_{2}$ crystal has a layered structure \cite{AdvPhys.18.193}. The Ta atomic layer is sandwiched between two S atomic layers. In the normal state, the Ta atoms sit at the center of approximate S octahedra. In the CDW phase, $1$\emph{T}-TaS$_{2}$ forms a $\sqrt{13}\times\sqrt{13}$ superstructure composed of the David star clusters, as shown in Fig.~\ref{david-star}(a). High-resolution STM image taken in this CDW phase is displayed in Fig.~\ref{david-star}(b). Close examination reveals the position of the topmost S atoms in each David star cluster. There exist three inequivalent types of Ta atoms in the unit cell designated by different colors in Fig.~\ref{david-star}(b) and (c).

The dI/dV spectrum, which is proportional to the electron local density of states (LDOS), is taken on the $1$\emph{T}-TaS$_{2}$ surface and plotted in Fig.~\ref{david-star}(d). The UHB and LHB can be clearly identified as the peaks marked in the dI/dV curve, in agreement with the previous reports \cite{PhysRevB.92.085132}. These spectroscopic features can be captured by DFT calculations including the on-site Hubbard U correction in a Hartree-Fock approximation \cite{PhysRevB.90.045134}. The peak below the LHB is referred to as the valence band (VB). The CDW gap thus locates between the LHB and the VB \cite{Smith_1985,PhysRevLett.81.1058,RevModPhys.78.17}. The conduction band (CB) is located above the UHB.

\begin{figure}
  \centering
  \includegraphics[scale=1.0]{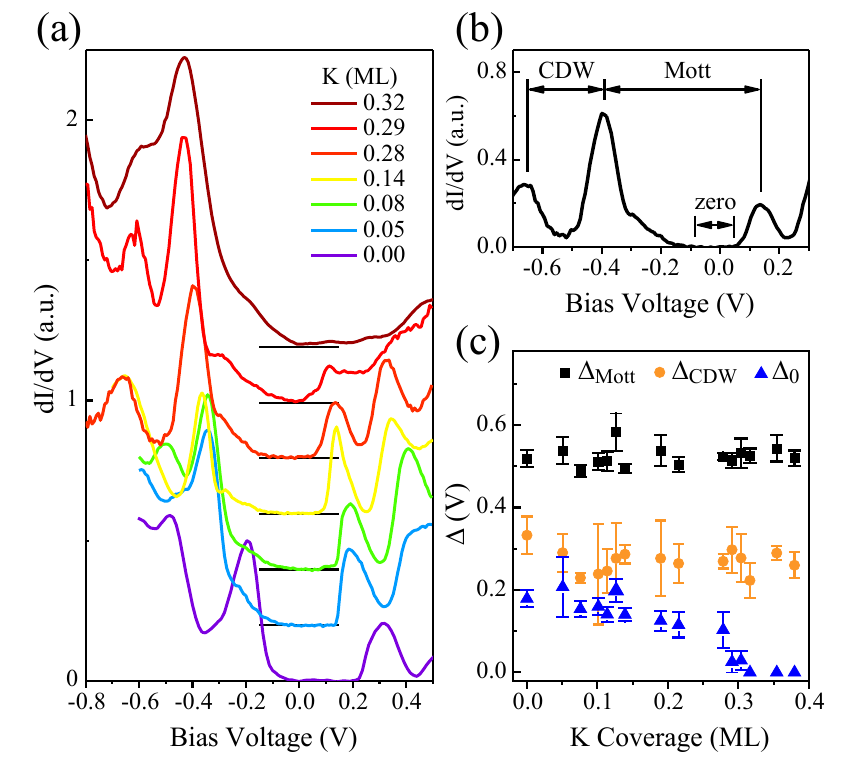}
  \caption{\label{coverage-dependence} (a) dI/dV spectra taken on the 1\emph{T}-TaS$_{2}$ surface with different K coverage. The data were acquired with $U=+500~\mathrm{mV}$ and $I_{t}=100~\mathrm{pA}$. (b) Definition of the three gaps in the $1$\emph{T}-TaS$_{2}$ dI/dV spectrum. $\Delta_{\mathrm{Mott}}$ is defined as the width between the two Hubbard peaks, $\Delta_{\mathrm{CDW}}$ the width between LHB and VB peaks, and $\Delta_{\mathrm{0}}$ the width of the zero-intensity region within the Mott gap. (c) Dependence of the magnitude of the three gaps on the K coverage.}
\end{figure}
The evolution of dI/dV spectra upon K doping to the 1\emph{T}-TaS$_{2}$ surface is plotted in Fig.~\ref{coverage-dependence}(a). Initially, the Fermi level starts to lift slightly towards the UHB. Yet, the whole characteristics of the spectra do not seem to change, as exemplified by the plot of 0.05 ML in Fig.~\ref{coverage-dependence}(a). This can be interpreted as that the K electrons are trapped by the surface impurities for a small amount of K deposition. When more K atoms are deposited, e.g., from 0.08 ML on in Fig.~\ref{coverage-dependence}(a), additional excitations become identifiable within the Mott gap, indicating that the K electrons dope the Mott insulator effectively.  An important feature is that these additional excitations are close to the LHB.
They keep growing until the whole Mott gap is filled up when the coverage reaches $\sim$0.32 ML. At this coverage, the collected dI/dV results  exhibit the uniform metallic state on the whole 1\emph{T}-TaS$_{2}$ surface, and thus the surface undergoes a transition from Mott insulator to metal. To quantitatively analyze the electronic evolution, we define the CDW gap $\Delta_{\mathrm{CDW}}$ as the width from the VB peak to the LHB peak, $\Delta_{\mathrm{Mott}}$ from the LHB peak to the UHB peak, and $\Delta_{0}$ for the zero intensity region accordingly, as illustrated in Fig.~\ref{coverage-dependence}(b). Figure~\ref{coverage-dependence}(c) shows the magnitude of the three gaps as a function of the K coverage. At first, $\Delta_{0}$ is closed at $\sim$0.32 ML. Secondly, during the Mott insulator-metal transition, the UHB and LHB peaks are both identifiable, and the magnitude of $\Delta_{\mathrm{Mott}}$ doesn't prominently change. Thirdly, no change is observed in the magnitude of $\Delta_{\mathrm{CDW}}$. The peak of VB remains at a fixed position relative to the LHB as the K coverage increases. Therefore, the electron doping by K atoms on the surface of $1$\emph{T}-TaS$_{2}$ doesn't close the CDW gap in the process of the metallization.

\begin{figure}
  \centering
  \includegraphics[scale=1.0]{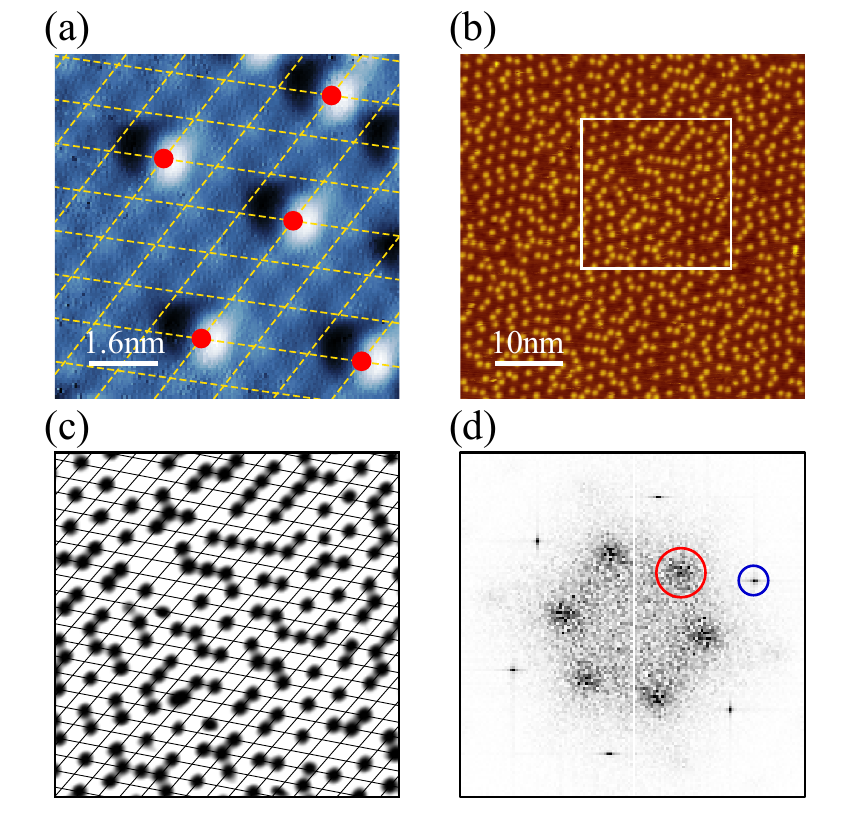}
  \caption{\label{cdw-coherence}(a) Small-scale STM current image ($8 \times8$ $\mathrm{nm}^{2}$) taken at the $1$\emph{T}-TaS$_{2}$ surface with a low K coverage. The David star structure is highlighted by the yellow dashed mesh, and the adsorption sites of K at the David star center are marked by red dots. (b) Large-scale STM topographic image ($50 \times50$ $\mathrm{nm}^{2}$) taken on the $1$\emph{T}-TaS$_{2}$ surface with the K coverage of $\sim$0.4 ML. $U=+1.0~\mathrm{V}$ and $I_{t}=100~\mathrm{pA}$. (c) The adsorption sites of K atoms extracted from the white square region in (b). The black dots represent the positions of K atoms, and the parallel lines the lattice of $1$\emph{T}-TaS$_{2}$ substrate. (d) Fourier transform of the STM image (b). The blue and red circles mark the peaks originated from the surface K atoms.}
\end{figure}
We further explore the effects of the alkali doping on the long-range CDW order.
Figure~\ref{cdw-coherence}(a) shows the STM topographic image of the $1$\emph{T}-TaS$_{2}$ surface with a low coverage of K atoms. The identified David star clusters form a triangular structure as traced by the yellow dashed mesh where each cluster is denoted by the crossing. The K atom is adsorbed preferentially on top of the center of the David star cluster. Then, we present the image of the surface covered with a high coverage of $\sim$0.4 ML in Fig.~\ref{cdw-coherence}(b), where all the K atoms are still in single-atom format without clustering. Figure 3(c) shows the zoom-in image of the white square region in Fig.~\ref{cdw-coherence}(b), where the black mesh and dots represent the David star structure and adsorbed K atoms, respectively.
We find that $152$ of the total $162$ K atoms are located at the David star center. This analysis suggests that the long-range CDW order still remains at this coverage. Moreover, Fig.~\ref{cdw-coherence}(d), the Fourier transform of Fig.~\ref{cdw-coherence}(b), shows that the peak corresponding to the CDW periodicity as marked by the blue circle is rather sharp, indicating the long-range CDW order without domain boundary. It is a remarkable result because the whole $1$\emph{T}-TaS$_{2}$ surface is already metallic at this coverage. In previous studies of the Mott insulator-metal transition in $1$\emph{T}-TaS$_{2}$, such as that induced by applying pressure \cite{nmat2318}, chemical substitution \cite{PhysRevX.7.041054}, ultra-fast laser pulses \cite{Stojchevska177}, current excitation \cite{ncomms11442,Yoshidae1500606} or applying a pulse voltage \cite{ncomms10956}, the domain boundaries were found to be responsible to the formation of the metallic state. However, in the K doped $1$\emph{T}-TaS$_{2}$ surface, the formation of domain boundary is not necessary for the metallization, and this Mott insulator-metal transition occurs without the destruction of the long-range CDW order.

\begin{figure}
  \centering
  \includegraphics[scale=1.0]{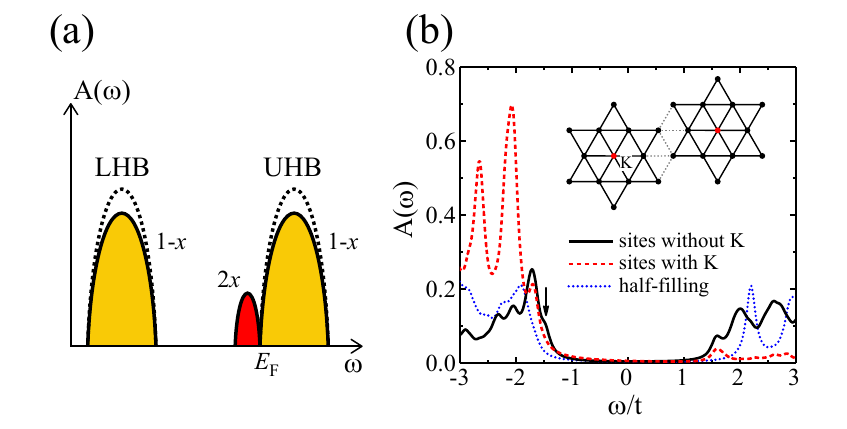}
  \caption{\label{theoritical-spectra}(a) Schematic of the spectral weight transfer in the local limit for an electron-doped Hubbard model with an uniform $U$. (b) Spectra calculated by the cluster perturbation theory. The spectrum of the half-filled system is illustrated by the dotted (blue) line, while the spectra in the sites with and without K$^{+}$ ions which are simulated by different $U$ (see the text) after one-electron doping in the $13$-site cluster are illustrated by the dashed (red) line and solid (black) line, respectively. The arrrow indicates the ingap excitation near the LHB. The inset shows the shape of clusters used in the theoretical calculations.}
\end{figure}
With the presence of the long-range CDW order in mind, we expect that a single-band Hubbard model would capture the main physics of the Mott insulator-metal transition, as the isolated narrow band still exists. Figure~\ref{theoritical-spectra}(a) schematically illustrates the spectral weight transfer in the conventional Hubbard model with electron doping \cite{PhysRevLett.67.1035,PhysRevB.90.245102}. In the large $U$ ($U/t\rightarrow\infty$) limit and with the half-filling, the spectral weight integrals of the LHB and UHB divided by the number of lattice sites are both equal to $1$. When extra electrons are introduced into this system, both the possibilities of creating holes (removing electrons from the LHB) and doubly occupancies (adding electrons into the UHB) are reduced by the doping concentration $x$. Thus, the total spectral weight of LHB and UHB are changed to $1-x$, and the reduced spectral weight shifts to the bottom of the UHB which is just below the Fermi level, as illustrated in Fig.~\ref{theoritical-spectra}(a). However, this picture clearly contradicts with the experimental results as shown in Fig.~\ref{coverage-dependence}, because the additional excitation is observed near the top of LHB and the spectral intensity of LHB is increased.

As mentioned above, the experimental results show that each K$^{+}$ ion is located at the center of one David star, so the electron concentration for the David stars with K$^{+}$ ions is larger than those without K$^{+}$ ions due to the Coulomb attraction of the positive charge of K$^{+}$, which is equivalent to the decrease of the effective repulsive interactions between electrons in the lattice sites with K$^{+}$ ions. Thus, we construct a site-dependent Hubbard model as described by
\begin{align}\label{model-h}
H=\sum_{<i,j>\sigma}t_{ij}(c_{i\sigma}^{\dag}c_{j\sigma}+\mathrm{h.c.})+\sum_{i}U_{i}n_{i\uparrow}n_{i\downarrow}
\end{align}
where $t_{ij}$ is the effective hopping between the nearest-neighbor David stars and $U_{i}$ the effective on-site Coulomb repulsion.
In theoretical calculations, we set $t_{ij}$ to be uniform $t_{ij}=t$, and $U_{i}=10t$ for David stars without K$^{+}$ deposition and $U_{i}=2t$ for David stars with K$^{+}$ deposition. We calculate the LDOS by using the cluster perturbation theory \cite{PhysRevB.90.245102,PhysRevLett.84.522,SupMat}. As shown in the inset of Fig.~\ref{theoritical-spectra}(b), to preserve the rotation symmetry of the triangular lattice, we use $13$-site clusters and set the K$^{+}$ ion at the center of each cluster. It is noteworthy that each site in the model represents a David star in the real material. The theoretical results of LDOS at half-filling and with one electron doping in the $13$-site clusters are shown in Fig.~\ref{theoritical-spectra}(b). For the half-filling system, a Mott gap of about $4t$ is clearly seen, while for the doped system, there exist additional excitations inside the Mott gap and close to the LHB for the David stars without K$^{+}$ deposition. This feature is qualitatively consistent with the experimental observation. We also present the LDOS for the David stars with K$^{+}$ deposition in Fig.~\ref{theoritical-spectra}(b). Since the effective interaction $U$ becomes small, most spectral weights are transferred to the low-energy part. The coupling between the David stars with and without K$^{+}$ deposition causes the redistribution of the spectral weight on the David stars without K$^{+}$ deposition, which leads to the appearance of the in-gap excitations near the LHB as observed in the experiment.

In summary, we realize a novel Mott insulator-metal transition in the surface of the CDW-assisted Mott insulator $1$\emph{T}-TaS$_{2}$ by surface alkali doping. Different from those triggered by the domain boundaries of the CDW state in which the long-range order of the CDW is destructed, we find that the long-range CDW order here is preserved all the way in the metallization, and the transition results from the filling up of the Mott gap by the in-gap additional excitations induced by the surface alkali doping. Moreover, the in-gap additional excitations are found to locate above the LHB, in contrast to the case of electron doped conventional Mott insulators where it is underneath the UHB. Making use of the numerical calculations of the site-dependent Hubbard model, we suggest that it is the K$^{+}$ ions adsorbed at the center of David stars that efficiently modulate the Coulomb energy and thus result in the appearance of the in-gap excitations near the LHB. This study paves the way toward the understanding of the correlation between CDW and Mott insulators in transition metal dichalcogenides.

\begin{acknowledgments}
This work was supported by the National Natural Science Foundation of China (Grants No. 11774149, 11790311, 11674158, 11774152), National Key Projects for Research and Development of China (2014CB921103, 2016YFA0300401).

X.Y.Z. and S. W. contributed equally to this work.
\end{acknowledgments}

~\\
\noindent \emph{Note added.}---This work is premised on the assumption that the insulating state in $1$\emph{T}-TaS$_{2}$ is due to Mott gap opening in the narrow band crossing the Fermi level, and the assumption that the physics in essentially 2D. There are other proposals that the metal-insulator transition is related to stacking order changes because of the strong interlayer coupling \cite{PhysRevLett.122.106404}.

\bibliography{1T-TaS2}

\end{document}


\title{Supplementary Material for: ``Realization of metallic state in $1$T-TaS$_{2}$ with persisting long-range coherence of charge density wave"}

\author{Xin-Yang Zhu}
\affiliation{National Laboratory of Solid State Microstructures and Department of Physics, Nanjing University, Nanjing 210093, China}
\author{Shi Wang}
\affiliation{National Laboratory of Solid State Microstructures and Department of Physics, Nanjing University, Nanjing 210093, China}
\author{Zhen-Yu Jia}
\affiliation{National Laboratory of Solid State Microstructures and Department of Physics, Nanjing University, Nanjing 210093, China}
\author{Li Zhu}
\affiliation{National Laboratory of Solid State Microstructures and Department of Physics, Nanjing University, Nanjing 210093, China}
\author{Qi-Yuan Li}
\affiliation{National Laboratory of Solid State Microstructures and Department of Physics, Nanjing University, Nanjing 210093, China}
\author{Wei-Min Zhao}
\affiliation{National Laboratory of Solid State Microstructures and Department of Physics, Nanjing University, Nanjing 210093, China}
\author{Cheng-Long Xue}
\affiliation{National Laboratory of Solid State Microstructures and Department of Physics, Nanjing University, Nanjing 210093, China}
\author{Yong-Jie Xu}
\affiliation{National Laboratory of Solid State Microstructures and Department of Physics, Nanjing University, Nanjing 210093, China}
\author{Zhen Ma}
\affiliation{National Laboratory of Solid State Microstructures and Department of Physics, Nanjing University, Nanjing 210093, China}
\author{Jinsheng Wen}
\affiliation{National Laboratory of Solid State Microstructures and Department of Physics, Nanjing University, Nanjing 210093, China}
\affiliation{Collaborative Innovation Center of Advanced Microstructures, Nanjing University, Nanjing 210093, China}
\author{Shun-Li Yu}
\affiliation{National Laboratory of Solid State Microstructures and Department of Physics, Nanjing University, Nanjing 210093, China}
\affiliation{Collaborative Innovation Center of Advanced Microstructures, Nanjing University, Nanjing 210093, China}
\author{Jian-Xin Li}
\affiliation{National Laboratory of Solid State Microstructures and Department of Physics, Nanjing University, Nanjing 210093, China}
\affiliation{Collaborative Innovation Center of Advanced Microstructures, Nanjing University, Nanjing 210093, China}
\author{Shao-Chun Li}
\affiliation{National Laboratory of Solid State Microstructures and Department of Physics, Nanjing University, Nanjing 210093, China}
\affiliation{Collaborative Innovation Center of Advanced Microstructures, Nanjing University, Nanjing 210093, China}

\date{\today}

\begin{abstract}
Here, we show the details of the cluster perturbation theory for the theoretical calculations of the single-particle spectral weight in the main text.
\end{abstract}

\maketitle

\begin{figure}
  \centering
  \includegraphics[scale=0.28]{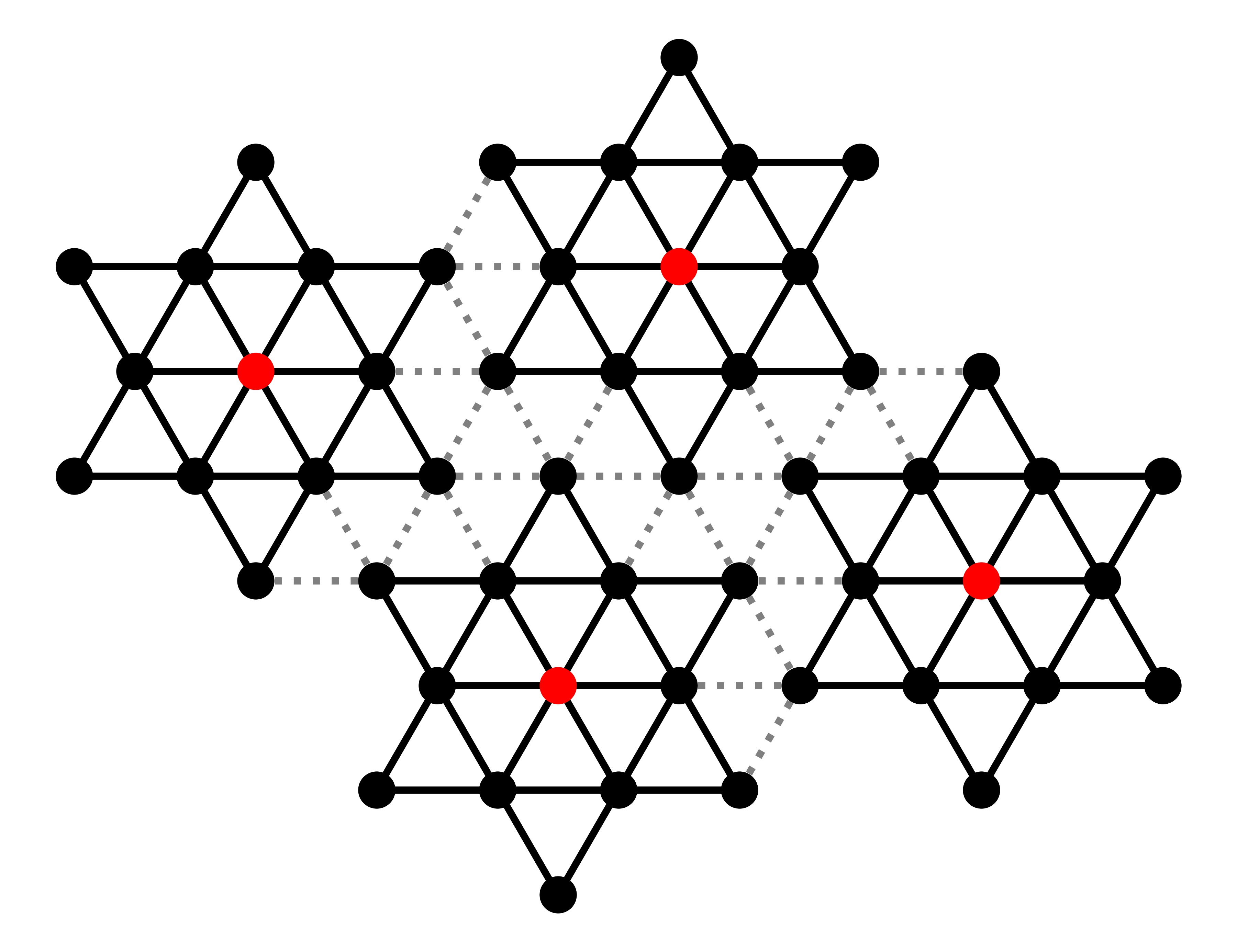}
  \caption{\label{cpt-cluster}Tiling of the triangular lattice with $13$-site clusters. The nearest-neighbor hoppings within a cluster are represented by solid lines, and intercluster hoppings by dashed lines. Here, each site represents a David star in $1$T-TaS$_{2}$, and the red sites are the David stars where the K$^{+}$ cations are located in the theoretical calculations.}
\end{figure}
In the main text, the computational technique for our calculation of the single-particle spectral weight $A_{i}(\omega)$ on site $i$ is based on the cluster perturbation theory (CPT), which has been successfully applied to many strongly correlated systems \cite{PhysRevLett.84.522,PhysRevLett.85.2585,PhysRevLett.92.126401,PhysRevLett.107.010401,PhysRevB.84.064520,PhysRevB.90.245102,PhysRevB.98.134410}. As shown in Fig.~\ref{cpt-cluster}, in the CPT calculations, we divide the original lattice into identical $13$-site clusters which constitute a superlattice. Then, the lattice Hamiltonian is written as
\begin{align}
H=H^{\prime}+V,
\end{align}
where $H^{\prime}$ is the cluster Hamiltonian, obtained by severing the hopping terms between different clusters, and $V$ contains the intercluster hoppings. The CPT method is based on a strong-coupling perturbation expansion of the one-body operators $V$ linking the individual clusters\cite{PhysRevLett.84.522}. At the lowest order of this expansion, the Green's function $\bm{G}$ of the original lattice can be expressed (in matrix form) as
\begin{align}
\bm{G}(\tilde{\bm{k}},\omega)=\bm{G}^{\prime}(\omega)[1-\bm{V}(\tilde{\bm{k}})\bm{G}^{\prime}(\omega)]^{-1},
\end{align}
where $\bm{G}^{\prime}$ is the cluster Green's function, and $\tilde{\bm{k}}$ is the wavevector in the Brillouin zone (BZ) of the superlattice. $\bm{G}^{\prime}$ is independent of $\tilde{\bm{k}}$, while $\bm{V}$ is frequency independent. The cluster Green's function $\bm{G}^{\prime}$ is calculated by the exact diagonalization method with the Lanczos algorithm, and it has the following expression,
\begin{align}
G^{\prime}_{\mu\nu}(\omega)&=\langle0|c_{\mu}\frac{1}{\omega-H+E_{0}+i\eta}c^{\dag}_{\nu}|0\rangle \nonumber \\
&+\langle0|c^{\dag}_{\nu}\frac{1}{\omega+H-E_{0}+i\eta}c_{\mu}|0\rangle,
\end{align}
where $E_{0}$ is the energy of the ground state $|0\rangle$, and $\eta$ is a small real number introduced in the calculation to shift the poles of the Green's function away from the real axis. Here, $\mu$ and $\nu$ denote both the site and spin degrees of freedom in a cluster.
The spectral function of the local single-particle excitation is given by
\begin{align}
A_{\mu\mu}(\omega)=-\frac{1}{\pi}\sum_{\tilde{\bm{k}}}G_{\mu\mu}(\tilde{\bm{k}},\omega).
\end{align}

\bibliography{TaS2-supmat}